\documentstyle[aps]{revtex}

\begin{document}
\title{Dynamic properties of solitons in the Frenkel-Kontorova Model. Application
to incommensurate CDW conductors\bigskip\ }
\author{Victor M. Burlakov}
\address{Institute of Spectroscopy Russian Academy of Sciences, Troitsk, Moscow\\
region, 142092, Russia}
\maketitle

\begin{abstract}
\bigskip\ 

An impact of kink-type solitons on infrared lattice vibrations is studied
for incommensurate Frenkel-Kontorova model. It is shown that the vibration
of particles involved into the kink formation is very similar to that in a
gap mode around the force constant defect. IR phonon mode intensity is found
to possess a universal dependence on the system parameters and the kink
concentration. It is argued that the giant IR peak observed in a number of
incommensurate charge density wave conductors can be explained in terms of
dynamic charge transfer stimulated by kinks. \medskip\ \ 

\bigskip
\end{abstract}

\section{Introduction}

A system of interacting particles in sinusoidal external potential
(Frenkel-Kontorova (FK) model \cite{frenk}) is widely used for description
of a broad variety of physical phenomena, such as statics and dynamics of
incommensurate phases (see, e.g. \cite{phas}), transport properties in
quasi-one dimensional conductors (see Ref. 3 and references therein),
adatoms diffusion on a metal surface \cite{adat}, etc. Peculiar features of
the FK model usually explored are related to the kink-like solitons.
Properties of the kinks have been described in a number of publications \cite
{kink1,kink2,kink3,kink4,kivsh1,kivsh2}. The dynamics of the FK model has
also been extensively studied, but mostly in relation to dynamics of
incommensurate superstructures rather than to the single kink \cite
{dyn1,dyn2,dyn3,dyn4}. Whereas it is not completely clear yet at what system
parameters the single-kink effects can be still important.

Incommensurate charge density wave (CDW) conductors are the physical systems
for which the single-kink effects can be very important. Vibration and
transport properties of these systems attract great interest due to numerous
peculiarities. The most striking among them are: i) the giant peak of
unknown origin in the low frequency infrared (IR) spectrum of a number of
inorganic CDW conductors such as $K_{0.3}MoO_3$, $(TaSe_4)_2I$ and $TaS_3$ 
\cite{don,kim}; ii) nonlinear conductivity and noise generation when
conducting dc current by these materials.

The aim of the present study is to investigate an impact of both single kink
and kink lattice onto IR active phonon spectrum and to specify the range of
the model parameters in which its properties can be treated in terms of
nearly independent kinks rather than in terms of superstructure, associated
with the kink lattice. I believe that after comparison of the numerical
results with the experimental data and proper justification of the model it
can serve as a bases for understanding of the microscopic nature of the
aforementioned peculiar features of CDWs.

The investigations were performed in two approaches: i) molecular dynamic
(MD) simulation was used for the system to reach an equilibrium state
according to the method proposed in Ref.\cite{aubry}, after what all the
particles were subjected to a small uniform step-like displacement and
subsequent vibrations were analyzed via Fourier-transformation; ii)
eigenvector problem (EVP) was solved in the harmonic approximation to study
the vibration spectrum of the system. The kinks in this case were taken into
account through expansion of the potential energy around particle
equilibrium positions obtained from MD simulation.

\section{Vibration spectrum in the presence of kinks}

Let us consider a chain of particles of mass $m$ and charge $e$ with nearest
neighbor interaction in the sinusoidal external potential $V\left( x\right)
=-\frac{V\cdot a^2}{4\pi ^2}\cdot \cos \left( 2\pi \cdot \frac xa\right) $
where $a$ is the potential period. In case of harmonic interparticle
interaction the motion equation for $n$-th particle is 
\begin{equation}
m\cdot \frac{\partial ^2U_n}{\partial t^2}+\gamma \cdot \frac{\partial U_n}{%
\partial t}+K_2\cdot (2U_n-U_{n-1}-U_{n+1})+\frac{V\cdot a}{2\pi }\cdot \sin
\left( 2\pi \cdot \frac{U_n}a\right) =e\cdot E(t),  \label{eq1}
\end{equation}
where $\gamma $ is phenomenological damping and $E(t)$ is external electric
field. Let the time dependent position $U_n$ of the particle can be
represented as $U_n(t)=n\cdot a+U_n^0+\delta _n(t),$where $U_n^0$ is
quasistatic variable describing a shift of the equilibrium position of the
particle with respect to the corresponding potential minimum, $\delta _n(t)$
describes a vibration of the particle around the new equilibrium position $%
U_n^0$. Then suggesting $\delta _n(t)=\delta _n(\omega )\cdot \exp (i\omega
\cdot t)$ and $E(t)=E_0\cdot \exp (i\omega \cdot t)$ the Eq. (\ref{eq1}) can
be splitted into two equations 
\begin{equation}
K_2\cdot (2U_n^0-U_{n-1}^0-U_{n+1}^0)+\frac{V\cdot a}{2\pi }\cdot \sin
\left( 2\pi \cdot U_n^0\right) =0,  \label{eq2}
\end{equation}

\begin{equation}
\delta _n(\omega )\cdot \left[ V\cdot \cos \left( 2\pi \cdot U_n^0\right)
-\omega ^2+i\omega \cdot \gamma \right] +\frac{K_2}m\cdot (2\delta _n(\omega
)-\delta _{n-1}(\omega )-\delta _{n+1}(\omega ))=\frac em\cdot E_0.
\label{eq3}
\end{equation}
Disregarding the trivial case $U_n^0=0$ when number of particles $N_{part}$
is equal to the number of potential minima $N_{pot}$, Eq. (\ref{eq2})
describes quasistatic kink-like deformation of the chain (due to neglection
of the dynamical term we restrict our consideration by standing kinks only).
Eq. (\ref{eq3}) describes the particle vibration around the new equilibrium
position. In the continues limit Eq. (\ref{eq2}) reduces to the sine-Gordon
equation \cite{SineGord} with the single-kink solution \cite{kinksol}: $%
U_n^0(i)=2a\cdot \pi ^{-1}\cdot arctg\left\{ \exp \left[ \pm \frac{2\cdot
(n-i)\cdot a}{R_k}\right] \right\} ,$ $R_k=2a\sqrt{\frac{K_2}V}$ can be
considered as the kink radius, $i$ is the kink position. Substituting this
solution into Eq. (\ref{eq3}) one can obtain the complex susceptibility $%
\chi (\omega )=\frac 1{E_0}\cdot \sum \delta _n(\omega )$, where the peaks
in $%
\mathop{\rm Im}
\left( \chi (\omega )\right) $ correspond to resonances $\omega _r$ and $%
\mathop{\rm Re}
\left( \delta _n(\omega _r)\right) $ corresponds to suitably normalized
eigenvector of the mode at $\omega _r$.

Fig.1 shows the $\omega (k)$ plot without (a) and with (b-c) kinks in the
model system. One can see the smearing of the resonances in the presence of
single kink (Fig.1a) while this smearing is gone in the system with the kink
lattice (Fig.1c). Instead, the phonon band folding due to the kink lattice
and the additional low frequency vibration arise. The latter is a so-called
phason, which is related to translational motion of kink(s) (domain
wall(s)). In case of negligible Peierls-Nabarro potential barrier the phason
frequency tends to zero. The phason is IR active and looks like a steep
increase at $\omega \longrightarrow 0$ in the optical conductivity spectrum $%
\sigma \left( \omega \right) =\omega \cdot 
\mathop{\rm Im}
\left( \chi (\omega )\right) $ (see Fig.2). The high frequency peaks in
Fig.2 correspond to phonons, the strongest one being related to in-phase
vibration of the particles situated near bottom of potential wells.

\section{Similarity between kink and the force constant defect}

The smearing of the vibrational states in the presence of single kink shown
in Fig.1b suggests that the kink is acting like a point defect. Moreover,
the particles involved into the kink formation are almost completely
eliminated from the high frequency phonon-like normal modes while these
particles obviously possess a higher vibration amplitude (local vibration
density) at low frequencies (see Fig.3). This is quantitatively illustrated
in Fig.4 where the eigenvectors of phason mode and that of IR phonon mode
are shown. Accordingly, one could try to describe vibration properties of
kinks in terms of localized vibrations around force constant defect. That
means the original incommensurate ($N_{part}\neq N_{pot}$) FK model is
replaced by commensurate one (or the ordinary harmonic chain of particles
with harmonic incite and interparticle potentials) and some particle cites
possess a defect force constant. The strength of the defect has been
determined from equation \cite{Barker}

\begin{equation}
1+\frac{\Delta V}N\cdot 
\mathop{\displaystyle \sum }
\limits_{k=\frac \pi N}^\pi \frac 1{V+4K_2\cdot \sin ^2\left( \frac k2%
\right) }=0,  \label{eq4}
\end{equation}
what means the zero-frequency gap mode formation in the vicinity of the
defect cite. As it is shown in Fig.4 the eigenvector of the gap mode is very
close to that of the phason while the eigenvectors of the phonon-like mode
nearly coincide in both cases. The corresponding spectrum of the 1D crystal
with the force constant defect is also shown in Fig.2. Note, that the
localization length of the gap mode $S_{gap}$ (the halfwidth of the peak
shown by dashed line in Fig.4) is equal to $R_k/\sqrt{2}$ in a wide range of 
$R_k$ values (see insert in Fig.4).

Thus, one may consider the system with kinks as a defect, or impurity
crystal taking for the description of its vibration properties all the
results already known. For instance, it is well understood that $S_{gap}$ is
determined basically by splitting of the gap mode from the bottom of the
optical band $\omega _0=\sqrt{V}$ and by the optical bandwidth $2\sqrt{K_2}$%
. One may argue therefore that the similarity between phason and the gap
mode eigenvectors and the phason eigenvector itself does not depend on the
potential anharmonicity provided that its influence on the above mentioned
parameters is small enough. It is thought therefore that the obtained
results will be applicable for a more realistic interparticle potential too.
From the analogy between the kinks and the defects it follows also that the
IR phonon mode intensity will show a linear decrease versus $n_k$ ($n_k=%
\frac{N_k}{N_{part}}$ is the kink concentration and $N_k=|N_{part}-N_{pot}|$
is the number of kinks) for low kink concentration. This indeed does take
place in certain range of $R_k$ values.

\section{When the kink effects are important}

Although the N-kink solution of Eq. (\ref{eq2}) is also available \cite
{nkinksol} it is more convenient to approximate it with the sum of
single-kink solutions. Our MD study of the ground state of a system
consisting of 128 particles arranged in 128-$N_k$ potential wells with
cyclic boundary conditions showed that even for $N_k\gg 1$ the kink lattice
can be perfectly described as a sum of the single-kink solutions with $%
R_k\simeq 2a\sqrt{\frac{K_2}V}$. Namely, for $N_k=8$ and $K_2=4*V$ ($%
R_k^{theor}=4.0a$) the value of $R_k^{\exp }$ $\simeq 3.94a$ has been
obtained. Similar results have been obtained also for the case when the
number of potential wells exceeds that of the particles. The dipole moment
spectrum $I\left( \omega \right) =%
\mathop{\rm Im}
\left[ \frac{\sum \delta _n(\omega )}{E_0}\right] $ has been both calculated
from (\ref{eq3}) substituting $U_n^0=\sum\limits_{i=1}^{N_k}U_n^0(i)$ with $%
R_k=R_k^{\exp }$ and obtained from MD simulation via fluctuation dissipative
approach for various values of $\eta =R_k\cdot n_k$. Both approaches agree
rather well even at very low frequencies although the harmonic approximation
obviously fails at $\omega =0$. Two examples of the particles arrangement
and corresponding eigenvectors of the IR vibrations are presented in Fig.5.
The eigenvectors for $\eta =0.25$ can be represented as a superposition of
the single-kink eigenvectors while for $\eta =0.5$ they are looking quite
different: even those particles which still occupy the potential wells and
are not involved into the kinks formation, are strongly involved into the
characteristic IR vibration (compare (a) and (b) in Fig.5). It should be
pointed out that there is no noticeable difference between the commensurate
and incommensurate cases (kink lattice period is equal and is not equal to
an integer number of $a$, respectively) if the kink concentration is not too
high. Otherwise the difference manifests itself in a small shift of the
zero-frequency peak in Fig.2 from its position.

I concentrate here on a question concerned with the intensity of the phonon
peaks in Fig.2 as a function of the parameter $\eta $. The EVP approach (Eq.
(\ref{eq3}) with various values of $\frac{K_2}V$ and $n_k$) was used for
this study. The results are presented in Fig.6. The integrated intensity $%
I_\Sigma =\int I\left( \omega \right) d\omega $ of the phonon peaks reveals
a universal dependence on the parameter $\eta $ \cite{burl}. It was also
found that the eigenvector of the strongest IR vibration obey some sort of
scaling invariance: the vectors obtained at different $n_k$ but for one and
the same $\eta $ can be transformed to each other by proper scaling of $a$.
Note, that the parameter $\eta $ means a volume fraction (in 1-D case)
occupied by the kinks and the observed decrease in $I_\Sigma $ at low $\eta $
values can be interpreted as washing out of the high frequency density of
states by gap modes associated with kinks. At higher $\eta $, when the kinks
form real lattice and eventually sinusoidal superstructure due to
interaction with each other, the decrease in $I_\Sigma \propto \eta $ slows
down because the real kink radius can not exceed at least one half of the
kink lattice period. Indeed, the linear decrease in $I_\Sigma $ shown in
Fig.6 ends at a cut-off value of $\eta \simeq 0.4$ (for $a=1$) which implies
the above mentioned restriction on the kink radius is $R_k\leq $ $0.4\cdot
k_s^{-1}$, $(k_s=n_k)$ where $k_s$ is the kink lattice (or superstructure)
wave vector measured in units of $\frac \pi a$. Thus, one can display a
range of parameters $k_s\cdot \sqrt{\frac{K_2}V}$ $\leq 0.2$ in which the
single-kink effects are important, i.e. the system can not be explicitly
treated in terms of some sinusoidal superstructure related to the kink
lattice. Since the IR eigenvectors has been argued to be not very sensitive
to anharmonicity one might expect this criterion holds for more realistic
potentials too.

\section{CDW collective dynamics}

The low frequency excitation spectrum of the CDW ground state have been
widely investigated both theoretically and experimentally (see, for example,
reviews \cite{rev1,rev2,rev3} and references therein). It is well
established that incommensurate CDW ground state is characterized by two
specific collective excitations: IR active phase mode and Raman active
amplitude mode.\cite{walker} The frequency of the former, $\omega _p,$ in
the incommensurate CDW conductors is of the order of $1cm^{-1}$ while the
amplitude mode frequency $\omega _a$ is about one-two orders of magnitude
higher. These vibrations have been observed experimentally in such model CDW
conductors as $K_{0.3}MoO_3$, $(TaSe_4)_2I$ and $TaS_3$ \cite
{kim,don,trav1,trav2,tom,sherwin,tkim,deg1}.

Besides the phase and the amplitude modes an additional vibration obeying
giant IR activity has been observed in all the above mentioned compounds 
\cite{don,kim}. The frequency of this additional feature in $(TaSe_4)_2I$ is
about $38cm^{-1}$ in between the phase ($\sim 1cm^{-1}\;$\cite{twkim}) and
the amplitude ($\sim 90cm^{-1}\;$\cite{tom}) mode frequencies. Several
explanations have been proposed to account for the additional giant IR peak,
but microscopic origin of this vibration is still not clear (see, for
instance, the discussions in Refs.1 and 15). In phenomenological model \cite
{deg2} the additional giant IR peak was thought to result from a bound
collective-mode resonance localized around impurity, but again without
emphasis of microscopic origin of the model parameters.

Below it is shown that the giant IR resonance occurs in the incommensurate
CDW system even in the absence of any impurities provided that the dynamical
charge transfer between adjacent CDW periods is taken into account and the
CDW possesses some kink lattice structure rather than sinusoidal structure.

Since the lattice deformation coupled to the CDW is much smaller than the
crystal lattice constant it is reasonable to describe the CDW system within
the FK model. The CDW periods in this case are associated with particles of
mass $m$ and charge $e$ which are placed into sinusoidal external (crystal
lattice) potential $V\left( x\right) =-\frac{V\cdot a^2}{4\pi ^2}\cdot \cos
\left( 2\pi \cdot \frac xa\right) $. The interparticle distance in the
commensurate phase is accepted to be equal to $2a$ what means the CDW is
formed due to dimerization. In case of $2N_{part}\neq N_{pot}$ one again
obtains incommensurate structure (kink lattice) and the time dependent
position $U_n$ of the particle can be represented as $U_n(t)=2\cdot n\cdot
a+U_n^0+\delta _n(t)$.

\subsection{Dynamic charge transfer}

As it has been demonstrated by Itkis and Brill\cite{brill}, spatial
redistribution of the charge condensed in CDW takes place under action of
static electric field. Obviously a characteristic time for the charge
redistribution or, in the other words, for the charge transfer from one CDW
period to another is determined by the amplitude mode frequency ($\sim
90cm^{-1}\;$\cite{tom}). Therefore, in the case of the giant peak frequency
the adiabatic condition is fulfilled.

To take into account the charge transfer contribution to the IR intensity of
any mode let us suppose that the particle charge in our model is determined
as 
\begin{equation}
\widetilde{e}_n(t)=e\cdot (1+\beta \cdot (\delta _{n+1}(t)-\delta
_{n-1}(t))),  \label{eq5}
\end{equation}
what means that the charge is transferred from the region of local
compression of the CDW to a region of local dilatation. The factor $\beta $
determines the fraction of the particle charge transferred during vibration.
The dipole moment $P(t)$ is determined as a sum of the part related to the
particles displacement $P_0(t)$ and the part, related to the charge transfer
between adjacent unit cells $P_{ct}(t)$ 
\begin{equation}
\begin{array}{c}
P(t)=P_0(t)+P_{ct}(t)= \\ 
\mathop{\displaystyle \sum }
\limits_ne\cdot \left[ \delta _n(t)+\beta \cdot \left(
U_n^0-U_{n-1}^0\right) \cdot (\delta _{n+1}(t)-\delta _{n-1}(t)-\delta
_n(t)+\delta _{n-2}(t))\right] .
\end{array}
\label{eq6}
\end{equation}
In commensurate phase $U_n^0-U_{n-1}^0=a$ for all $n$ and the charge
transfer dipole moment $P_{ct}(t)$ vanishes according to (\ref{eq6}). In
incommensurate phase the $P_{ct}(t)$ value can be rather high. It will be
shown that the charge transfer effect is essentially determined by
kink-related disturbance of the periodicity in the particle arrangement.
Using the criterion of importance of single-kink effects obtained in the
preceding section one can examine if the kinks are important for description
of the charge density wave conductor $(TaSe_4)_2I$. The superstructure wave
vector in this system is $k_s\simeq 0.085$ \cite{wavevec}, $\sqrt{V/m}$ can
be associated with the giant IR peak frequency $\omega \sim 0.005\,eV$ \cite
{rich} and $\sqrt{K_2}$ can be estimated from the phason dispersion $\sqrt{%
K_2/m}<0.001\,eV$ \cite{phas_disp}. Thus, one obtains $k_s\cdot \sqrt{\frac{%
K_2}V}\ll 1$ what implies that the kink effects can be important in this
compound.

Fig.7a shows the fragment of MD simulation of arrangement of 128 particles
over 264 potential wells, i.e. the CDW with superstructure. The 51-th
particle (shown by arrow in the figure) is pinned. The conductivity spectra $%
I\left( \omega \right) =\omega \cdot 
\mathop{\rm Im}
\left[ \frac{\sum \delta _n(\omega )}{E_0}\right] $ obtained from MD
simulation are shown in Fig.8 for both pinned and depinned system. The
features of the interest are the phase mode (PM) and the peak of CT mode
(charge transfer mode), marked by PM- and CT-arrows in Fig.8, respectively.
The latter peak is genetically related to the vibration with the wave vector
equal to that of the superstructure. The corresponding eigenvectors are
shown in Fig.7b. Taking into account that the CDW internal deformation can
be adiabatically accompanied by charge redistribution one obtains that the
CT peak acquires the giant IR intensity. The conductivity spectra in which
the charge transfer effect has been taken into account according to Eqs (\ref
{eq5}) and (\ref{eq6}) are shown in Fig.8 by symbols (depinned chain) and
thin solid line (pinned chain). The phase mode intensity is almost
independent on the charge transfer effect while the CT mode intensity
increases several orders in magnitude.

Fig.9 shows that the CT mode intensity $\left( 
\mathop{\rm Im}
\left[ \frac{\sum \delta _n(\omega )}{E_0}\right] \right) $ with charge
transfer contribution decreases with the increase of the parameter $4a\sqrt{%
\frac{K_2}V}$ and possesses a universal dependence regardless the
superstructure wave vector. Physically this feature can be understood in the
following way. The charge transfer dipole moment consists of a sum of
elementary dipole moments resulting from the charge transfer over the
inter-kink distance. The longer is the latter the higher is the elementary
dipole moment, but the smaller is the number of these dipole moments. Thus,
the total charge transfer dipole moment does not depend on the kink
concentration (probably unless $4a\sqrt{\frac{K_2}V}<a/n_k$). On the other
hand, this dipole moment strongly depends on the kink-mediated distortion of
the chain. The latter decreases with the increase of the kink radius
resulting in the observed decrease in CT mode intensity in Fig.9.

It was experimentally proved that the giant IR peak intensity in $%
(TaSe_4)_2I $ increases with the increase of the sample temperature \cite
{physb}. This unusual feature can be naturally explained in terms of the
charge transfer effect discussed above. Indeed, as it is clear from Fig.9
the integrated intensity of the CT mode as a function of the system
parameters can be approximated as (see dashed line in Fig.9) 
\begin{equation}
I\simeq C_0\cdot e^2\cdot \left[ \sqrt{\frac{K_2}V}\right] ^{-2.25},
\label{eq7}
\end{equation}
where $C_0$ is constant. The interparticle force constant, obviously,
depends on the particle charge $K_2=e^2K_2^{^{\prime }}$ since the particles
are associated with the charges condensed in CDW. The (\ref{eq7}) can be
then rewritten as 
\begin{equation}
I\simeq C_0\cdot e^{-0.25}\cdot \left[ \sqrt{\frac{K_2^{^{\prime }}}V}%
\right] ^{-2.25}.  \label{eq8}
\end{equation}
Thus, despite the decrease of the particle charge (the CDW amplitude) the CT
mode integrated intensity increases upon approaching $T_p\simeq 261K$ from
below! Due to short range order the particle charge (CDW amplitude) remains
finite even at very high temperatures and accounts for the high CT mode
intensity well above $T_p$.

Note one more peculiarity of the CT peak. It does not depend on the number
of particles in the coherent CDW domain or, in the other words, on the
effective mass of CDW condensate. The latter can explain why the
corresponding frequency has nearly the same value in such different
compounds as $K_{0.3}MoO_3$ and $(TaSe_4)_2I$ \cite{deg1,twkim}.

\section{Summary}

The kink-like solitons in the incommensurate Frenkel-Kontorova model are
investigated regarding to their impact on the vibration spectrum. It is
found that the IR phonon intensity possesses universal dependence on the
product of the kink radius and the kink concentration suggesting some sort
of scaling invariance for the corresponding eigenvector. The model
accounting for the giant IR peak in the incommensurate inorganic CDW
conductors is proposed. It is shown the giant IR peak is related to the
fundamental vibration with the wave vector equal to that of the
superstructure and the giant IR intensity is caused by dynamical charge
transfer accompanying the CDW internal motion.

\section{Acknowledgments}

This work was partially supported by Russian Ministry of Science through the
program ''Fundamental Spectroscopy''.

\begin{center}
\newpage\ FIGURE CAPTURES
\end{center}

Fig.1. Dispersion of vibration band obtained via EVP solution (system of
linear equations (\ref{eq3})) for the FK model containing 64 particles
arranged over: (a) 64 potential wells (no kinks); (b) 63 potential wells
(one kink); (c) 56 potential wells (four kinks). The dark regions correspond
to the higher vibration amplitude of particles excited by external field $%
E=E_0\cos (k\cdot n)\cdot \cos (\omega t)$.

Fig.2. Conductivity spectra of the FK model with one kink. (1) is the
calculated spectrum; (3) is that obtained by MD simulation (32 particles
with cyclic boundary conditions) for $K_2=4\cdot V,$ $\sqrt{V}=72$ $arb.un$;
(2) is the spectrum corresponding to the force constant defect $\Delta
V=-4.1231\cdot V$ (see text), in the position of the particle no.16.

Fig.3. MD study of local density of states distribution over the particles
in the commensurate (a) and containing one kink (b) FK model. The particles
were initially subjected to random displacements and the temporal evolution
of the spatial harmonics was analysed via Fourier-transformation. The dark
regions correspond to the maxima in the Fourier spectrum.

Fig.4. The kink and the phonon (the strongest peak in Fig.1) eigenvectors
obtained as described in the caption to Fig.1. By symbols in the insert is
shown the dependence of the gap mode radius upon kink radius $R_k=2\sqrt{%
\frac{K_2}V}$.

Fig.5. (a) The particle arrangement in the FK model of 128 particles (shown
by symbols) in 120 potential wells (solid line). (b) The eigenvectors of the
kink-(1,3) and phonon-like (2,4) modes. $K_2=4\cdot V$ (solid symbols) and $%
K_2=16\cdot V$ (open symbols), $\sqrt{V}=72$ $arb.un$.

Fig.6. Integrated intensity of the phonon-like modes upon $\eta =R_k\cdot
n_k $ calculated using (\ref{eq4}) for FK model of 128 particles arranged
over: (1) 112 potential wells ($n_k=1/8$); (2) 120 potential wells ($%
n_k=1/16 $); and (3) 124 potential wells ($n_k=1/32$).

Fig.7. (a) Fragment of particle arrangement obtained via MD simulation in
the FK model containing 128 particles arranged over 272 potential wells ($%
n_k=1/16$, see text) for $K_2=16\cdot V$. The particle No.51 is pinned by an
extra local potential. (b) Eigenvectors of phason mode without pinning
(dotted line) and with pinning (thick solid line), and those of CT-mode (see
Fig.2) shown by thin solid line for both pinned and depinned chain.

Fig.8. IR conductivity spectra of the FK model (see caption to Fig.7),
calculated using Eqs. (3)-(5) for $4\sqrt{\frac{K_2}V}=8$ ($a=1$). Thin
solid line is for $\beta =0$ and thick solid line is for $\beta =30$ in case
of pinned chain. Dashed line is for $\beta =0$ and symbols are for $\beta
=30 $ in case of depinned chain. PM are the phase modes and CT are the modes
which intensity may contain the charge transfer contribution. For $\beta =30$
the $0.03\cdot e$ of the particle charge is transferred during the CT mode
vibration while for $\beta =0$ it is $0$.

Fig.9. Dependence of the CT mode integrated intensity on the potential
parameters of the FK model containing 128 particles arranged over 264 minima
($n_k=1/32$): (1) is for $\beta =30$ and (4) is for $\beta =0$; over 272
minima ($n_k=1/16$): (2) is for $\beta =30$ and (5) is for $\beta =0$. (3)
is the dependence $y=0.45/x^{2.25}$. Arrow shows the parameters for the
spectra presented in Fig.8.

\end{document}